%% file: astroph.tex
\documentclass[11pt, preprint]{aastex}

\shorttitle{A Spitzer White Dwarf Infrared Survey}
\shortauthors{Mullally et al.}

\begin{document}

\title{A Spitzer White Dwarf Infrared Survey}
\author{F. Mullally\altaffilmark{1},  
Mukremin Kilic\altaffilmark{1},
William T. Reach\altaffilmark{2},
Marc J. Kuchner\altaffilmark{3},
Ted von Hippel\altaffilmark{1},
Adam Burrows\altaffilmark{4} \and
D. E. Winget\altaffilmark{1}}
\altaffiltext{1}{Department of Astronomy, 1 University Station, C1400, Austin, TX 78712}
\email{fergal@astro.as.utexas.edu}

\altaffiltext{2}{ MS 220-6, California Institute of Technology, Pasadena, CA 91125}
\altaffiltext{3}{NASA Goddard Space Flight Center, Greenbelt, MD 20771}
\altaffiltext{4}{Department of Astronomy and Steward Observatory, University of Arizona, 933 North Cherry Avenue, Tucson, AZ 85721}

\begin{abstract}
We present mid-infrared photometry of 124 white dwarf stars with Spitzer Space Telescope. Objects were observed simultaneously at 4.5 and 8.0$\mu$m with sensitivities better than 1~mJy. This data can be used to test models of white dwarf atmospheres in a new wavelength regime, as well as to search for planetary companions and debris disks. 

\end{abstract}


\section{Introduction}

White dwarf stars (WDs) are the evolutionary end point of stellar evolution for all main sequence stars with a mass $\lesssim 8 M_{\odot}$ \citep{Weidemann00}. The mass of an isolated white dwarf is believed to be uniquely determined by the progenitor mass, hence the progenitor lifetime for a white dwarf can be estimated. Nuclear burning has ceased, so their evolution is one of monotonic and predictable cooling. From the mass and temperature of a white dwarf its cooling age can be calculated. A white dwarf is a stellar gravestone with a date of birth and death carved upon it.

Previous white dwarf infrared surveys have concentrated on the near infrared. \citet{Zuckerman92} surveyed 200 stars down to K=16 while \citet{Farihi05} observed 261 nearby WDs with limiting magnitude in J of between 18 and 21.
In the mid-infrared \citet{Chary98} surveyed 12 WDs with ISOCAM at 7 and 15$\mu$m. Our survey significantly extends these previous works by looking at a large (124) sample of stars at 4.5 and 8.0$\mu$m with a limiting sensitivity of better than 0.1mJy. This dataset allows us to study the behavior of white dwarf atmospheres in this wavelength range and to search for companion planets and disks. Such a large dataset will also undoubtedly be useful to other researchers for unanticipated reasons.

The primary purpose of our survey was to search for the presence of planets and brown dwarf companions. With radii $\sim 1~R_{\earth}$, WDs are orders of magnitude less luminous than their progenitor stars. This dramatically reduces the contrast between the host star and any orbiting daughter planets. \citet{Becklin88} reported on the first infrared search for substellar companions around WDs while \citet{Burleigh02} suggested using near infrared imaging to detect $\gtrsim 3M_J$ planets in orbits $> 5$~AU with 8m telescopes. Other attempts to directly detect a companion to a white dwarf include \citet{Debes05, Farihi05} and \citet{Debes06}. Theoretical spectra of brown dwarf stars and massive planets show a distinctive bump around 4-5$\mu$m between absorption bands of methane and water \citep{Sudarsky03, Burrows03}. By comparing the observed flux in this passband with that of a nearby passband we can hope to directly detect the companion as an excess to the white dwarf flux. 

Excess mid-infrared flux around a white dwarf can also be caused by a warm disk of circumstellar material. Fortunately, the spectral signature of a disk is markedly different from that of a planet or brown dwarf, showing a mostly flat continuum over a broad wavelength range. Prior to this work, only one white dwarf (G~29-38), included in our survey, was known to exhibit an infrared excess caused by such a disk \citep{Zuckerman87}. \citet{Reach05b} detected emission features of silicates at 10$\mu$m from this disk using a Spitzer IRS spectrum \citep{Houck04}. 

von Hippel et al. (2006, submitted, hereafter HKK06) proposed that debris disks may be the source metals observed in the photospheres of approximately 25\% of WDs and suggested that debris disks are therefore very common. Other surveys have increased the number of WDs with measured near infrared excesses consistent with disks to 5 \citep[][and references therein]{Kilic06}, including WD2115-560, a 9700~K hydrogen atmosphere (DA) white dwarf discovered as part of this survey (HKK06). Mid-infrared observations are sensitive to cooler dust at larger orbital separations and will be important in determining the orgin and lifetimes of these disks.

Our survey has also uncovered some unusual behavior of the SEDs of cool WDs. \citet{Kilic06} published an SED of WD0038-226 showing a dramatic flux deficit and noted that DAs below 7,000~K consistently showed a small flux deficit compared to that expected from blackbody models. These results provide an opportunity to investigate the properties of matter in extreme conditions, but uncertainty in the infrared luminosity of the coolest WDs is an obstacle to their use in white dwarf cosmochronology to measure the age of the Galaxy.

\section{Target Selection, Observations and Reductions}
Drawing from the McCook \& Sion Catalogue \citep{McCook99}, we cross-referenced with the 2MASS survey \citep{Skrutskie06}, selecting all stars brighter than $K_s$ = 15, rejecting known binaries and planetary nebulae, for a total of 135 objects. We removed one object to avoid conflict with the Reserved Observations Catalogue, and the Spitzer TAC removed 3 other WDs awarded to a different program. In total we observed 131 objects and successfully measured the flux for 124 of these. The remaining objects were too heavily blended with other, brighter objects. A breakdown of the spectral type of each object is given in Table~\ref{spt}. In the course of a more detailed literature search we discovered, for a small number of stars, differences between the temperature and spectral type quoted in McCook \& Sion compared to more recently published values. Where applicable, references are listed in the notes to Table~\ref{results}.

Each object was observed simultaneously in Channels 2 and 4 (4.5 and 8.0$\mu$m) with the IRAC camera \citep{Fazio04} on the Spitzer space telescope \citep{Werner04}. Five 30 second exposures were taken of each object using a Gaussian dither pattern. The data were processed with version S11.4.0 of the IRAC pipeline to produce the Basic Calibration Data (BCD) files which removes well understood instrumental signatures.

We performed aperture photometry on these BCD files using the astrolib package in IDL. For most stars, we chose a 5 pixel aperture, although for a number of objects we used 2 or 3 pixels instead to avoid contaminating flux from nearby objects. We measured sky in an annulus of 10-20 pixels centered on the star. We made the appropriate aperture correction suggested by IRAC data handbook. For channel 2 we multiplied the flux by 1.221, 1.113 and 1.050 for apertures of 2, 3 and 5, while the values for channel 4 were 1.571, 1.218 and 1.068 respectively. 

The recorded flux for a stellar object is dependent on the location on the array where it was observed. This is because of both a variation in pixel solid angle (due to distortion) and a variation of the spectral response (due to varying filter response with incidence angle over the wide field of view). We accounted  for these effects by multiplying the measured flux by the appropriate location dependent correction factor as described in \citet{Reach05a}. We do not apply a correction to our photometry to account for variation in the flux as a function of location of the stellar centroid within the pixel, as this correction only applies to data taken in Channel 1.

Because the sensitivity of the IRAC sensors is wavelength dependent, the recorded flux differs from the true flux in a manner that depends on the source's spectral shape. Fortunately this effect is small (of order the systematic uncertainty) and, as a white dwarf spectrum is dominated by a Rayleigh-Jeans tail in the mid-infrared, easily corrected. We used the values suggested in \citet{Reach05a} of 1.011 at 4.5$\mu$m and 1.034 at 8$\mu$m. We did not apply color correction to objects whose SEDs were inconsistent with a single blackbody source (see Table~\ref{results}). The IRAC pipeline removes some but not all cosmic rays. To clean our data of remaining artifacts we removed frames where the flux deviated by more than 3.5$\sigma$ from the median, and calculated the weighted average flux of the remaining frames.

\section{Results}
We present the fluxes for each white dwarf in the two IRAC bands in Table~\ref{results}. For comparison, we also list the flux for each object in J, H and K as measured by the 2MASS survey. An SED for each object, with optical photometry from the McCook \& Sion catalogue is presented in Figure 1. A blackbody at the quoted temperature and fit to the optical and near infrared data is also shown to guide the eye. A subset of objects in this sample have been previously published in \citet{Reach05b} and \citet{Kilic06}; we present photometry for these stars here for completeness. Note that as these papers used an earlier version of the IRAC pipeline their published fluxes differ slightly from those presented here.

\subsection{Notes on Individual Objects}
{\em WD0002+729} 
A handful of stars show a small excess at 8~$\mu$m; we discuss this star as an example object. The atmosphere of WD0002+729 is contaminated with small amounts of metals \citep{Wolff02} which increases the probability of the existence of a disk (HKK06). However, the flux is close to our sensitivity limits (approx 0.1mJy) and our error bar may be underestimated; our confidence in this excess is low. By fitting models to this excess we can determine that if this excess is due to a disk, its maximum temperature must be less than about 300~K.

{\em WD0031-274} McCook \& Sion incorrectly classified this object as a DA. \citet{Kilkenny88} classified it to be an sdOB star with their criterion ``dominated by HeI and HeII lines; often Balmer absorption present''. \citet{Lisker05} measures a temperature of 36,097~K and a distance of 900~pc. They refer to it as an sdB. Our photometry shows a clear excess from J onward relative to the visible photometry. Close examination of the images does not reveal any irregularities in the point response function. At this distance, the flux from a substellar companion would be negligible. We fail to find a good fit for a low mass main sequence star nor does the excess show the broad flat shape of a circumstellar dust spectrum. The IR data is best fit with a blackbody temperature of 18,300~K, however a circumstellar object of this temperature would be detectable in the visible flux. Cyclotron emission has been suggested as a source of infrared emission in WDs, but this object is not known to have a strong magnetic field. Further study is necessary to determine the true nature of this object.

{\em WD0038-226} See \citet{Kilic06} for a further discussion of this object's dramatic flux deficit.

{\em WD0447+176} McCook \& Sion, quoting \citet{Wegner90} gives a V magnitude of 13.4 ($\approx$ 11~mJy) and a temperature of 9,044~K. This magnitude is inconsistent with the other available photometry. Instead we use V=12.62 from \citet{Kilkenny88}, who measure a temperature of 33.8~kK. Neither of these temperatures fit the photometry well. We plot instead a best fit blackbody temperature of 15.5~kK.

{\em WD0843+358} This object partially resolves into 2 objects separated by 2.2 pixels at 8$\mu$m, causing the observed excess at that band. The companion object is not seen at any bluer wavelengths. Were the companion substellar in nature the excess would be greater at 4.5$\mu$m than at 8; we therefore conclude that the excess is due to a background object. Examination of POSS 1 plates from 1953 shows no evidence of an object at the current position of the WD.

{\em WD1036+433} Also known as Feige~34, this star is incorrectly listed in the McCook \& Sion catalogue as 36~kK DA. \citet{Thejll91} determine it be an 80~kK sdO. \citet{Maxted00} note that they observed H$\alpha$ in emission from this object, but the emission is intermittent as other observers make no mention of it \citep{Oke90,Bohlin01}. \citet{Chu01}, who also observe emission, suggest this emission could be caused by photoionisaton of the atmosphere of a hot Jupiter companion.

Our photometry shows a clear excess from H onward. Examination of the individual images shows that the white dwarf is the brightest object in the vicinity, and no evidence of a line of sight companion. This excess was first noticed in J,H and K by \citet{Probst83}. \citet{Thejll91} summarized the available photometry at that time and concluded that the colors were ``marginally consistent'' with a companion K7-M0 dwarf. We find that the excess is well fit in color and magnitude by a 3,750~K Kurucz model \citep{Kurucz79}, corresponding to a spectral type of M0 or M1.

{\em WD1234+481} \citet{Liebert05} measure a temperature for this star of 55,040~K $\pm$ 975, $\log{g}$ of 7.78 $\pm$ 0.06 and derive a distance of 144~pc. \citet{Holberg98} measure 56,400~K, 7.67 and 129~pc respectively based on an IUE spectrum. Other authors measure similar values. \citet{Debes05} noticed an excess in the near infrared, and assigned a preliminary spectral type to the companion of M8 V.

We observe an infrared excess in all 5 bands. The IRAC images are round and isolated. The measured flux values in apertures of 2, 3 and 5 pixels yield consistent values, ruling out the possibility that the excess is caused by contamination from a nearby bright star. We note that the excess can be fit by a model of a brown dwarf with $T_{eff} < 2000~K$ corresponding to a spectral type of early L. If planned follow-up observations confirm the sub-stellar nature of this companion it would be the fourth white dwarf -- brown dwarf binary known.

{\em WD1616-390} \citet{Sion88} list this star as a $0.61M_{\odot}$ DA with $T_{eff}$ of 24,007~K. We notice a clear excess from J onward. The colors of the excess are well fit with a 4000~K Kurucz model of a main sequence star, corresponding to a late K spectral type. However the magnitude of the excess is too large to be consistent with K dwarf companion. We conclude that the excess is either from a foreground dwarf star or a background giant. 

{\em WD2115-560} This object has an infrared excess consistent with a dust debris disk. See HKK06 for further details on this object.

{\em WD2134+218} This object was too faint to be detected at 8$\mu$m. The expected flux according to a blackbody model was 0.05mJy, less than our nominal detection limit of about 0.1mJy with the 150s exposure time used.

{\em WD2326+049} This object, also known as G29-38, has an infrared excess consistent with circumstellar dust. This excess was first reported by \citet{Zuckerman87}. \citet{Reach05b} fit a Spitzer IRS spectrum of the disk with a mixture of olivine, fosterite and carbon dust.

\section{Discussion}
We fit the observed SEDs of DA stars between 6-60~kK in the optical and
near infrared with synthetic photometry derived from models kindly
supplied by Detlev Koester. Details of the input physics and methods are 
described in  \citet{Finley97}, \citet{Homeier98} and \citet{Koester01}. We then compared the observed excess (or deficit) over the fitted model in the mid-infrared to the uncertainty in the observation. Objects with disks (WD2326, WD2115), probable companions (e.g WD1234)
or contaminated photometry are not included. We expect the distribution of this value for the sample to be well described by a Gaussian distribution with mean zero and standard deviation of one. We plot this distribution in Figure~\ref{histogram}. The grey histogram corresponds to IRAC Channel 4 photometry and the fitted gaussian is plotted with a dashed line. The outlined histogram corresponds to Channel 2 and the gaussian is shown as a solid line. The dispersion in Channel 2 is measurably greater than expected, indicating that our error bars may be underestimated. The measured flux in Channel 4 is on average 1 $\sigma$ higher than expected. This may be because the majority of stars have excess flux in this band (possibly due to a disk), or a poor match of our models to reality. As these models are well tested only in the optical regime for which they were originally intended, it is possible this result points to new and unexpected physical processes affecting the mid-infrared portion of the spectrum.

Three brown dwarf companions to WDs are known \citep{Becklin88,Farihi04, Maxted06}. In our survey we find one object with an SED consistent with a brown dwarf companion, or a detection frequency of approximately 1\%. \citet{Farihi05} surveyed 261 WDs and found no new brown dwarf companions setting the brown dwarf companion fraction of $\lesssim 0.5$\%, consistent with our result. Our selection of targets deliberately excluded stars with known main sequence binary companions explaining the dearth of such companions in our sample. Both subdwarf stars in our survey show an infrared excess. This is consistent with the survey of \citet{Allard94} who determined that 54-66\% of subdwarf stars have a main sequence companion.

We discovered one new object with a notable infrared excess consistent with a debris disk (WD2115-560), and re-observed a second (WD2326+049). HKK06 suggests that metals observed in the photosphere of some WDs are caused by accretion from debris disks and not from the ISM as suggested by others \citep{Dupuis93, Koester06}. Table~\ref{daz} lists the 22 objects in our survey with measured abundances of metal in their photospheres. The small fraction of DAZs with disks initially appears to refute this claim. It should be noted that for stars with temperatures $\gtrsim 19$kK, dust from any debris disk inside the Roche limit is expected to sublimate quickly and won't produce a noticeable infrared signature. For the cooler stars, the metal abundances are significantly lower and the flux from the debris disk is expected to be correspondingly dimmer. Deeper observations will be required to confirm or refute this hypothesis. 

\section{Conclusion}
We have conducted the first large mid-infrared white dwarf photometric survey, obtaining images of 124 WDs with a limiting sensitivity better than 0.1mJy. This survey has already found an unexplained flux deficit in the SED of a cool white dwarf, and a debris disk around another star. This dataset can be used to constrain the presence of planets around these stars as well as test and refine model white dwarf atmospheres.

\acknowledgments
This work is based on observations made with the Spitzer
Space Telescope, which is operated by the Jet Propulsion Laboratory,
California Institute of Technology under NASA contract 1407. Support for
this work was provided by NASA through award project NBR 1269551 issued
by JPL/Caltech to the University of Texas. This work is performed in part under contract with the Jet Propulsion Laboratory (JPL) funded by NASA through the Michelson Fellowship Program. JPL is managed for NASA by the California Institute of Technology. This publication makes use of data products from the Two Micron All Sky Survey, which is a joint project of the University of Massachusetts and the Infrared Processing and Analysis Center/California Institute of Technology, funded by the National Aeronautics and Space Administration and the National Science Foundation. This research has also made use of the SIMBAD database, operated at CDS, Strasbourg, France

{\it Facilities:} \facility{Spitzer (IRAC)}



\input{tab1}
\input{tab2}

\input{tab3}

\clearpage



\notetoeditor{Note to typesetter. This paragraph is the caption to the figure. The figure is comes as 7 files, one per page.}

\figcaption{Spectral energy distributions for the 124 white dwarf stars in this survey. Our IRAC photometry points are at 4.5 and 8$\mu$m. Also shown for comparison are near infrared fluxes from 2MASS, and B and V photometry from the McCook \& Sion catalogue where available. Points with arrows are upper limits on the flux, not measurements (see Table~\ref{results}). The solid line is a black body at the temperature listed in Table~\ref{results} fit to the optical and near IR data. A blackbody is a good, but not perfect model of a white dwarf photosphere, and this accounts for much of the deviation in the photometry. A better fit can be achieved with atmosphere models, however as these fits are intended only to guide the eye a blackbody does an adequate job.}

\clearpage
\begin{figure}
\plotone{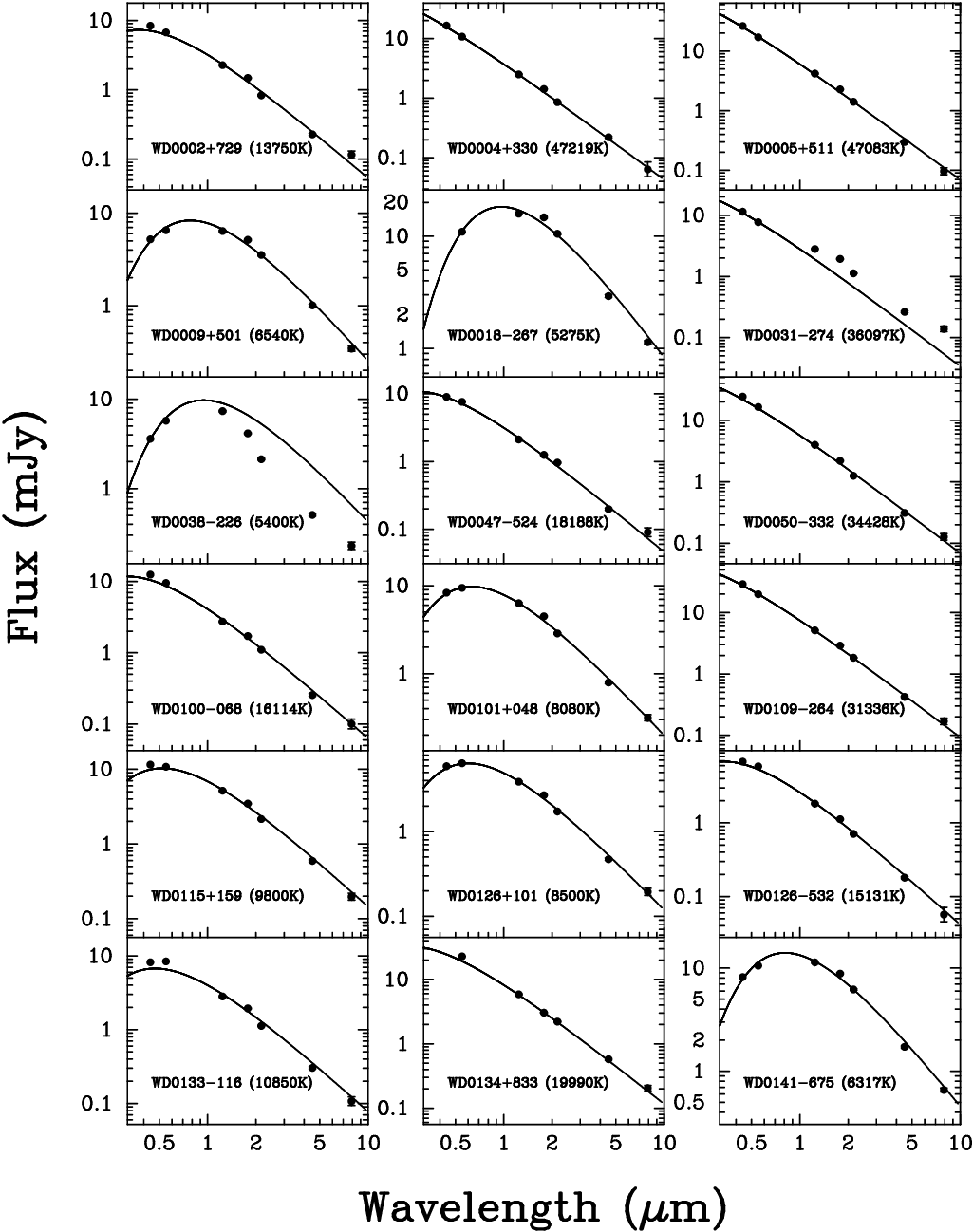}
\centerline{f01a.ps}
\end{figure}
\clearpage

\begin{figure}
\plotone{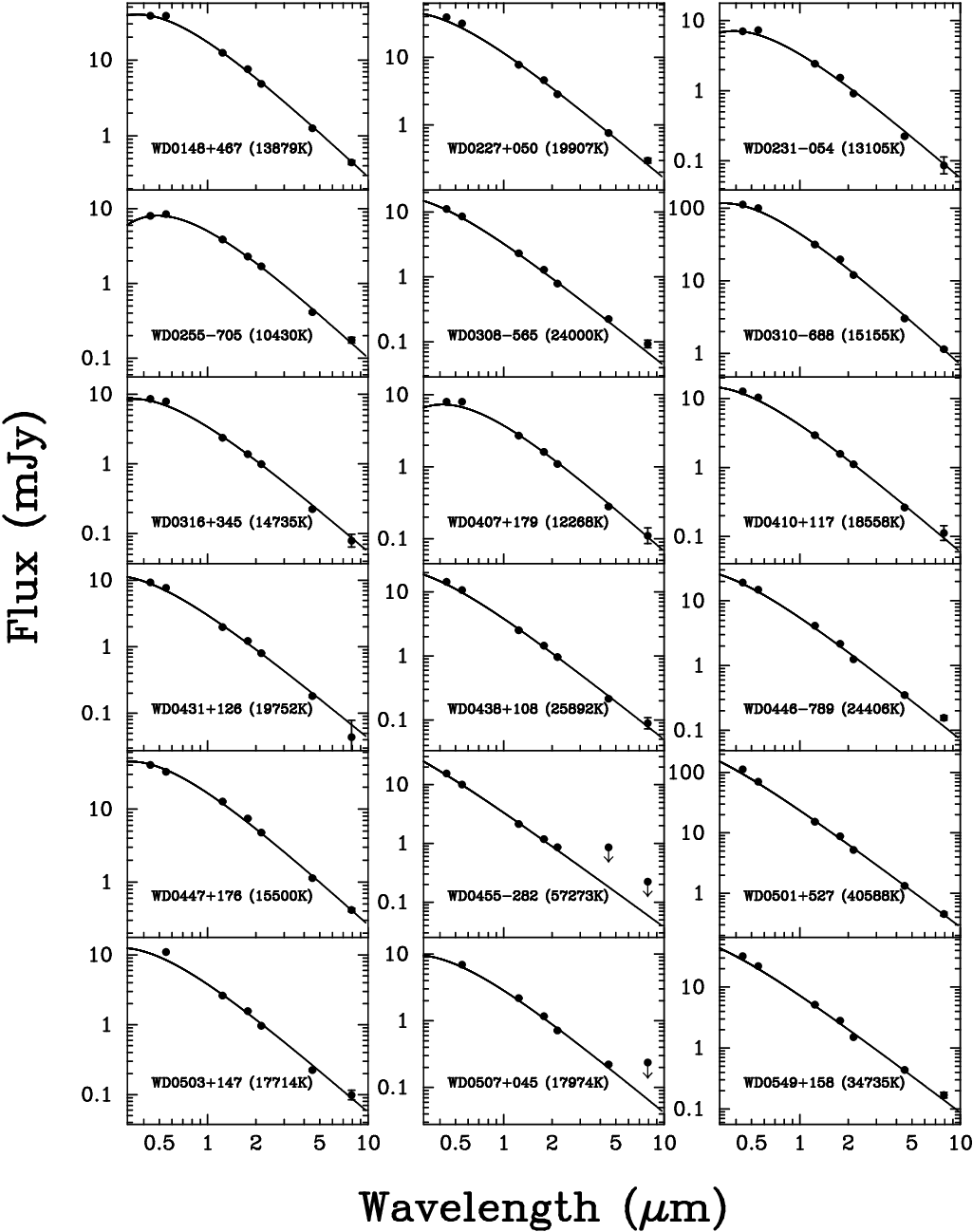}
\centerline{f01b.ps}
\end{figure}
\clearpage

\begin{figure}
\plotone{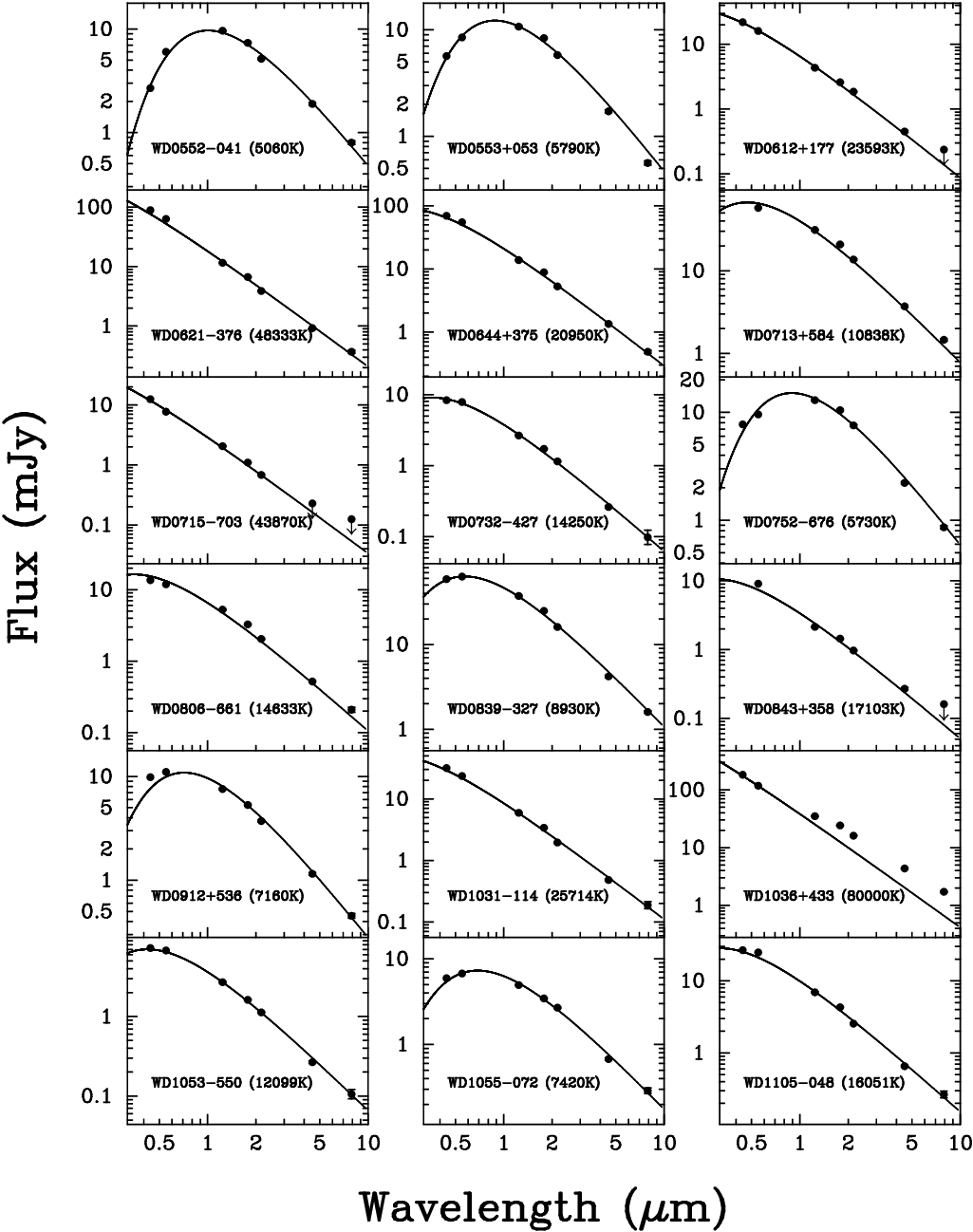}
\centerline{f01c.ps}
\end{figure}
\clearpage

\begin{figure}
\plotone{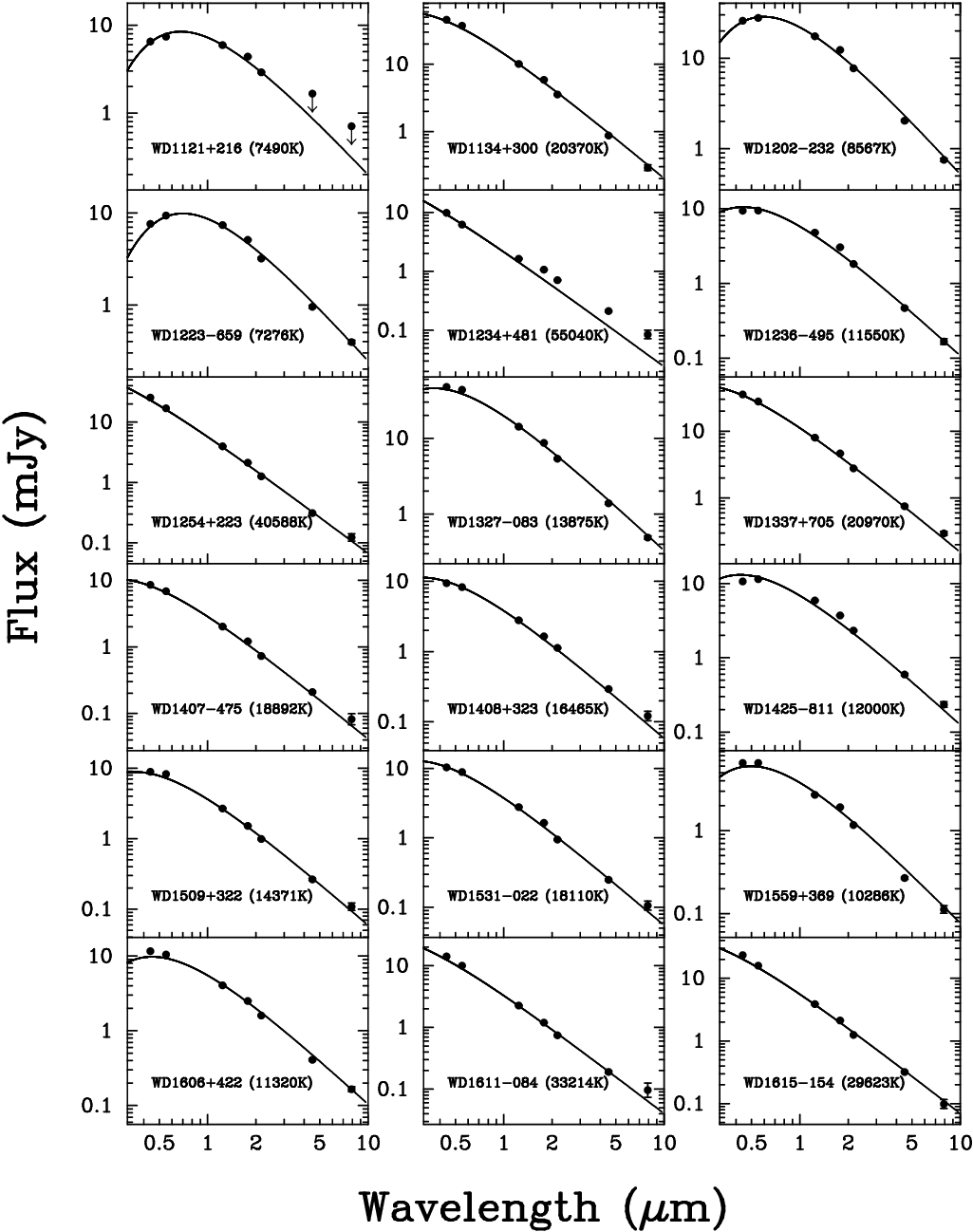}
\centerline{f01d.ps}
\end{figure}
\clearpage

\begin{figure}
\plotone{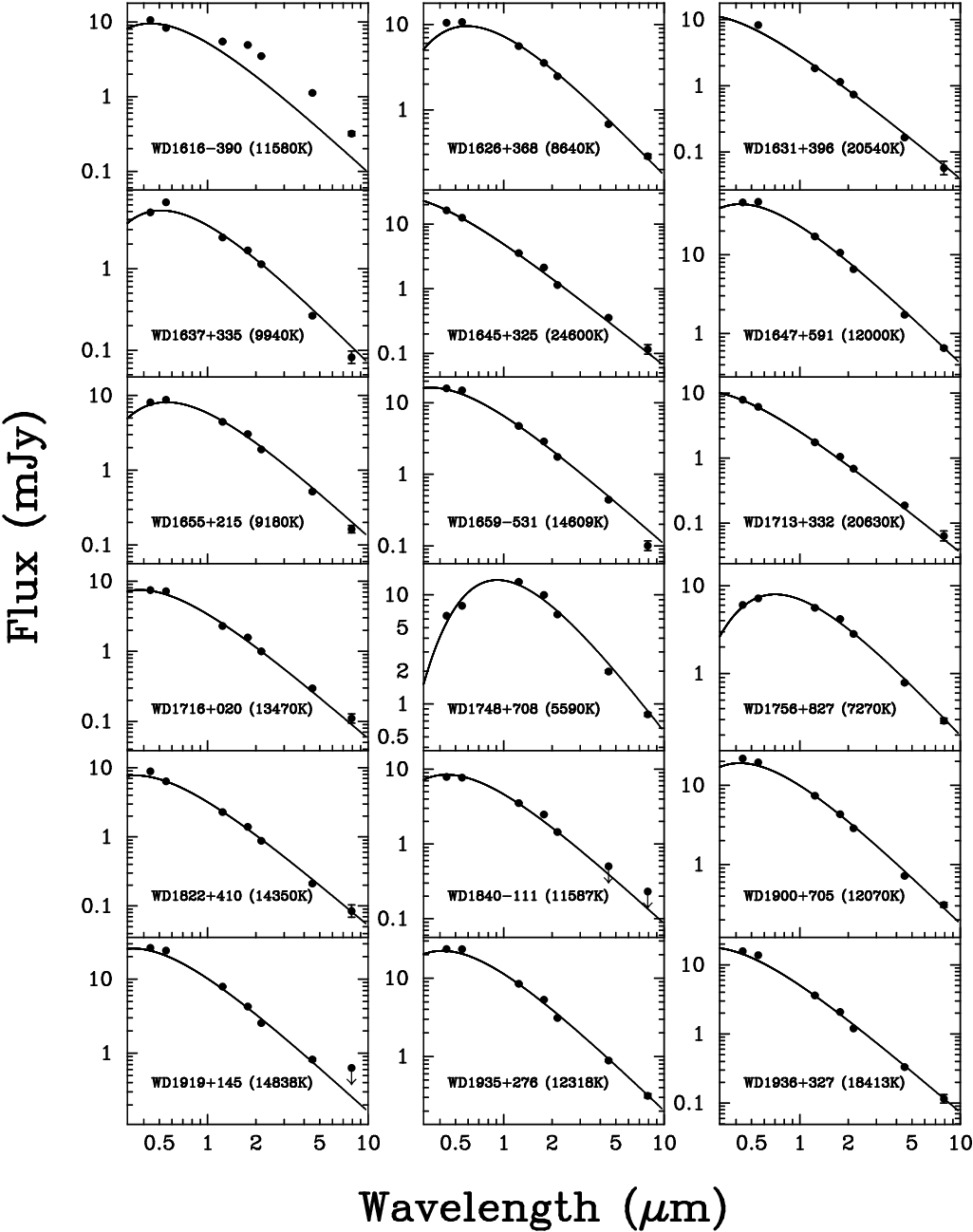}
\centerline{f01e.ps}
\end{figure}
\clearpage

\begin{figure}
\plotone{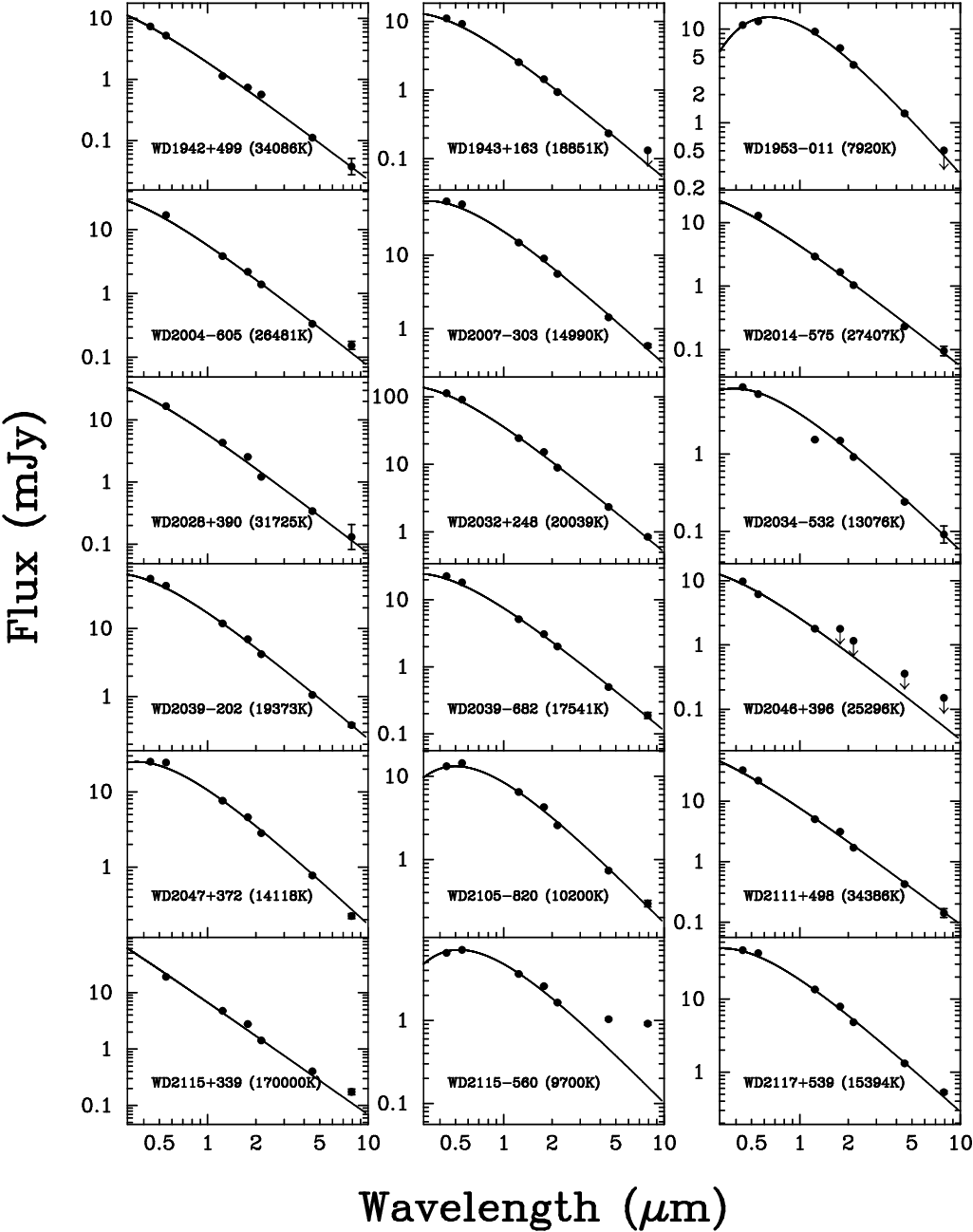}
\centerline{f01f.ps}
\end{figure}
\clearpage

\begin{figure}
\plotone{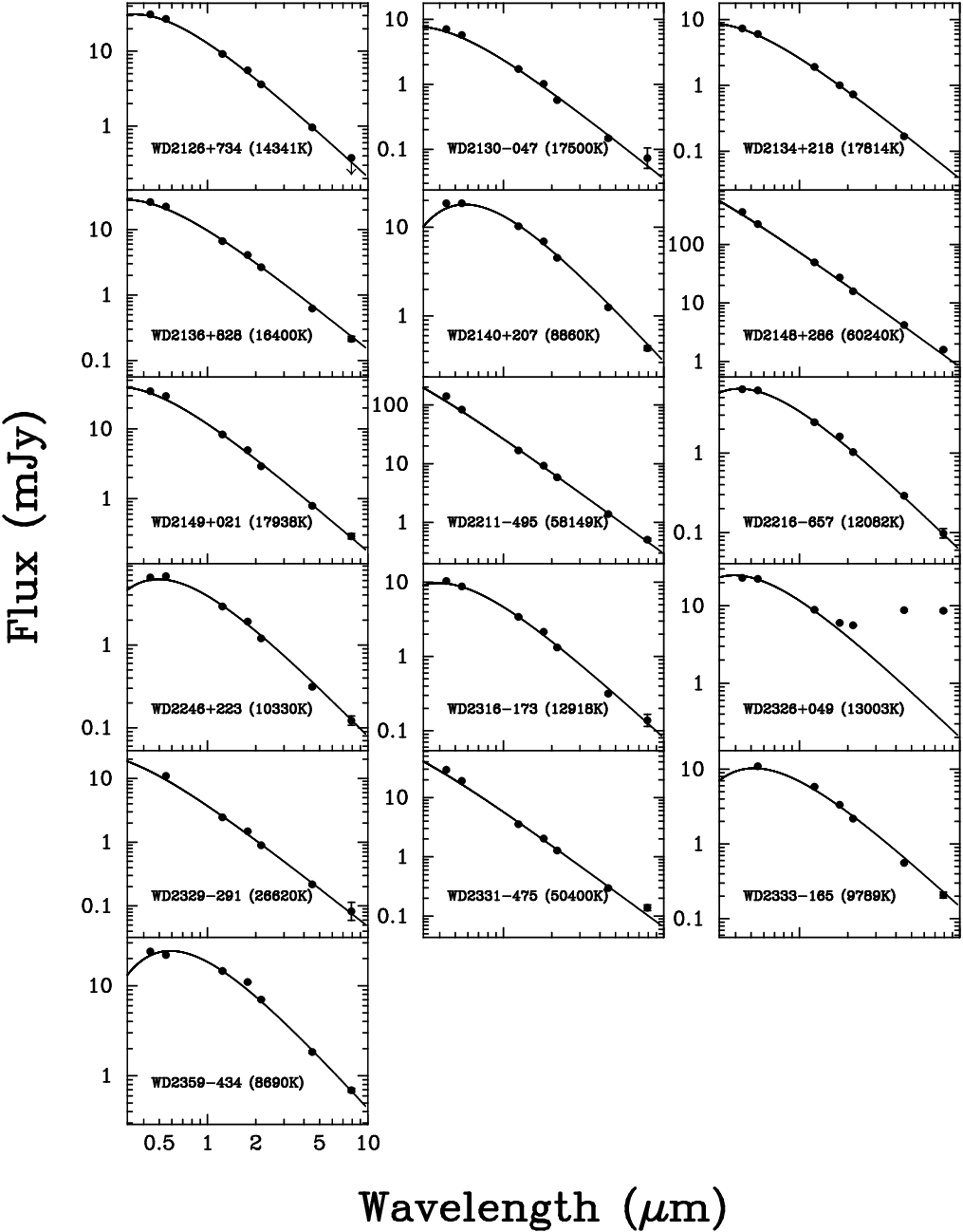}
\centerline{f01g.ps}
\end{figure}

\begin{figure}
    \includegraphics[scale=.75,angle=270]{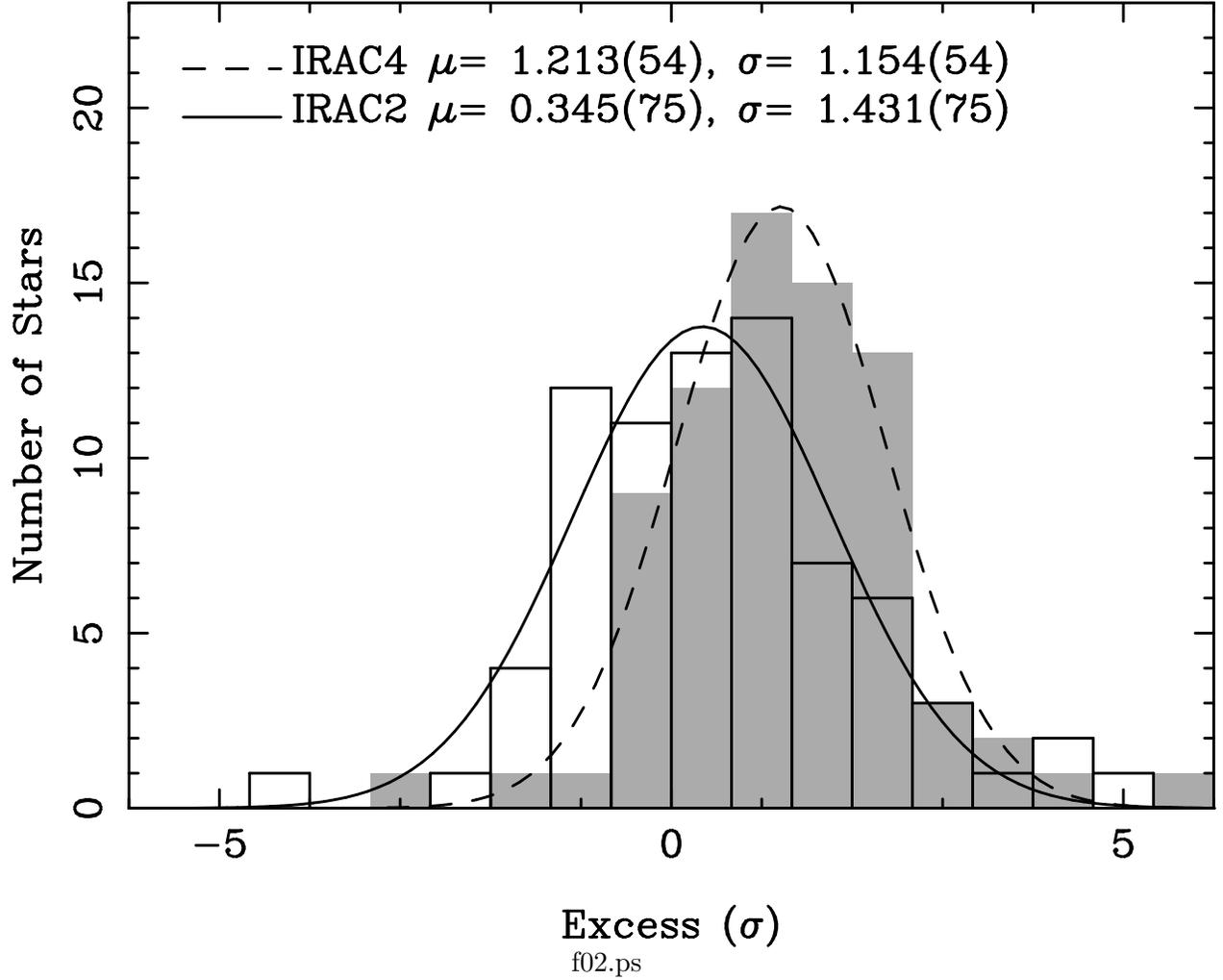}
    \centerline{f02.ps}
    \figcaption{Histogram of excesses and deficits for a selection of the DAs observed in this survey. Each bin represents the difference between the observed flux and that predicted by models from \citet{Finley97} as a fraction of the size of the measured photometric error. If no stars show unusual behavior, we expect this histogram to be well fit by a gaussian with mean zero and a standard deviation of 1. IRAC 2 is shown as white bars fit with the solid line while IRAC 4 is the grey histogram fit with the dashed line. The mean of each distribution is given by $\mu$ while the standard deviation is given by $\sigma$. See text for discussion \label{histogram}}
\end{figure}


\end{document}

%% file: tab1.tex
%

\begin{deluxetable}{lr}
\tablewidth{0pt}
\tablecaption{Classifications of Observed Stars\label{spt}}

\tablehead{
    \colhead{Type}&
    \colhead{Number}\\
}

\startdata
DA & 98\\
DB & 10\\
DC & 3\\
DO & 2\\
DQ & 2\\
sd & 2\\
Other & 7\\

\enddata
\end{deluxetable}

%% file: tab2.tex

\begin{deluxetable}{llrrrrrrccc}
\tabletypesize{\footnotesize}
\tablewidth{0pt}
\tablecaption{Infrared Fluxes for Stars in this Sample \label{results}}

\tablehead{
	\colhead{Name}&
	\colhead{Type}&
	\colhead{$T_{eff}$}&
	\colhead{J}&
	\colhead{H}&
	\colhead{K}&
	\colhead{IRAC 2}&
	\colhead{IRAC 4 }&
	\colhead{Ap}&
	\colhead{Notes}\\
}

\startdata

\object{WD0002$+$729} & DBZ & 13750 & 2.272(40) & 1.484(29) & 0.833(16) & 0.2277(79) & 0.116(16) & 3 & 1\\
\object{WD0004$+$330} & DA & 47219 & 2.503(44) & 1.416(28) & 0.857(16) & 0.2212(86) & 0.064(21) & 5\\
\object{WD0005$+$511} & DO & 47083 & 4.200(73) & 2.271(44) & 1.411(27) & 0.2975(99) & 0.096(14) & 3\\
\object{WD0009$+$501} & DAP & 6540 & 6.41(11) & 5.14(10) & 3.528(67) & 1.006(31) & 0.343(23) & 5 & 2\\
\object{WD0018$-$267} & DA & 5275 & 15.88(28) & 14.72(29) & 10.51(20) & 2.923(88) & 1.134(42) & 5\\
\object{WD0031$-$274} & sdB & 36097 & 2.816(49) & 1.937(38) & 1.124(21) & 0.2634(89) & 0.138(16) & 3 & 3\\
\object{WD0038$-$226} & $C_2H$ & 5400 & 7.34(13) & 4.141(81) & 2.132(40) & 0.507(16) & 0.229(23) & 5 & 2,4\\
\object{WD0047$-$524} & DA & 18188 & 2.117(37) & 1.257(25) & 0.961(18) & 0.1982(73) & 0.091(15) & 3\\
\object{WD0050$-$332} & DA & 34428 & 3.989(70) & 2.191(43) & 1.249(24) & 0.312(10) & 0.127(17) & 3\\
\object{WD0100$-$068} & DB & 16114 & 2.742(48) & 1.710(33) & 1.102(21) & 0.256(11) & 0.100(17) & 3\\
\object{WD0101$+$048} & DA & 8080 & 6.32(11) & 4.486(88) & 2.862(54) & 0.792(25) & 0.315(26) & 5\\
\object{WD0109$-$264} & DA & 31336 & 5.125(89) & 2.899(57) & 1.835(35) & 0.422(14) & 0.168(22) & 5\\
\object{WD0115$+$159} & DC & 9800 & 5.149(90) & 3.454(68) & 2.155(41) & 0.594(19) & 0.197(22) & 3\\
\object{WD0126$+$101} & DA & 8500 & 3.888(68) & 2.688(52) & 1.731(33) & 0.472(15) & 0.194(20) & 3\\
\object{WD0126$-$532} & DA & 15131 & 1.828(32) & 1.122(22) & 0.710(13) & 0.1801(66) & 0.057(14) & 3\\
\object{WD0133$-$116} & DAV & 10850 & 2.816(49) & 1.937(38) & 1.124(21) & 0.303(10) & 0.107(16) & 3\\
\object{WD0134$+$833} & DA & 19990 & 5.88(10) & 3.058(60) & 2.212(42) & 0.576(19) & 0.205(21) & 5\\
\object{WD0141$-$675} & DA & 6317 & 11.37(20) & 8.85(17) & 6.20(12) & 1.722(52) & 0.658(29) & 5 & 5\\
\object{WD0148$+$467} & DA & 13879 & 12.45(22) & 7.58(15) & 4.848(92) & 1.257(39) & 0.444(29) & 5\\
\object{WD0227$+$050} & DA & 19907 & 7.76(14) & 4.608(90) & 2.844(54) & 0.764(24) & 0.297(27) & 5\\
\object{WD0231$-$054} & DA & 13105 & 2.435(42) & 1.539(30) & 0.913(17) & 0.2249(87) & 0.086(27) & 5\\
\object{WD0255$-$705} & DA & 10430 & 3.873(68) & 2.292(45) & 1.695(32) & 0.412(13) & 0.174(16) & 3 & 6\\
\object{WD0308$-$565} & DB & 24000 & 2.308(40) & 1.286(25) & 0.787(15) & 0.2243(80) & 0.093(14) & 3\\
\object{WD0310$-$688} & DA & 15155 & 31.57(55) & 19.69(38) & 12.01(23) & 3.030(92) & 1.142(41) & 5\\
\object{WD0316$+$345} & DA & 14735 & 2.375(41) & 1.381(27) & 0.993(19) & 0.2234(80) & 0.079(18) & 3\\
\object{WD0407$+$179} & DA & 12268 & 2.709(47) & 1.610(31) & 1.097(21) & 0.280(10) & 0.110(31) & 5\\
\object{WD0410$+$117} & DA & 18558 & 2.941(51) & 1.577(31) & 1.115(21) & 0.2626(97) & 0.112(32) & 5\\
\object{WD0431$+$126} & DA & 19752 & 1.972(34) & 1.227(24) & 0.801(15) & 0.1815(76) & 0.044(34) & 5\\
\object{WD0438$+$108} & DA & 25892 & 2.522(44) & 1.448(28) & 0.962(18) & 0.2143(76) & 0.089(20) & 3\\
\object{WD0446$-$789} & DA & 24406 & 4.135(72) & 2.173(42) & 1.252(24) & 0.350(12) & 0.156(14) & 3\\
\object{WD0447$+$176} & DB & 15500 & 12.71(22) & 7.43(14) & 4.764(90) & 1.135(35) & 0.415(24) & 3\\
\object{WD0455$-$282} & DA & 57273 & 2.134(37) & 1.181(23) & 0.863(16) & 0.857(27) & 0.223(26) & 5 & 7\\
\object{WD0501$+$527} & DA & 40588 & 15.32(27) & 8.76(17) & 5.228(99) & 1.327(41) & 0.454(32) & 5\\
\object{WD0503$+$147} & DB & 17714 & 2.614(46) & 1.569(31) & 0.966(18) & 0.2247(77) & 0.099(17) & 2\\
\object{WD0507$+$045} & DA & 17974 & 2.180(38) & 1.167(23) & 0.714(14) & 0.2213(76) & 0.236(20) & 2 & 7\\
\object{WD0549$+$158} & DA & 34735 & 5.144(90) & 2.818(55) & 1.506(28) & 0.437(14) & 0.169(19) & 2\\
\object{WD0552$-$041} & DZ & 5060 & 9.63(17) & 7.35(14) & 5.166(98) & 1.892(58) & 0.799(34) & 5\\
\object{WD0553$+$053} & DAP & 5790 & 10.73(19) & 8.36(16) & 5.79(11) & 1.717(52) & 0.561(26) & 3\\
\object{WD0612$+$177} & DA & 23593 & 4.342(76) & 2.594(51) & 1.853(35) & 0.450(15) & 0.236(24) & 2 & 8,5\\
\object{WD0621$-$376} & DA & 48333 & 11.56(20) & 6.68(13) & 3.879(73) & 0.922(29) & 0.370(23) & 5\\
\object{WD0644$+$375} & DA & 20950 & 13.79(24) & 8.82(17) & 5.271(100) & 1.350(42) & 0.487(30) & 5\\
\object{WD0713$+$584} & DA & 10838 & 31.14(54) & 20.75(40) & 13.66(26) & 3.71(11) & 1.461(50) & 5\\
\object{WD0715$-$703} & DA & 43870 & 2.055(36) & 1.095(21) & 0.683(13) & 0.2281(93) & 0.124(22) & 5 & 9\\
\object{WD0732$-$427} & DAE & 14250 & 2.660(46) & 1.720(34) & 1.148(22) & 0.2620(92) & 0.098(25) & 3\\
\object{WD0752$-$676} & DAZ & 5730 & 12.94(23) & 10.47(20) & 7.57(14) & 2.219(67) & 0.859(30) & 3 & 2\\
\object{WD0806$-$661} & DQ & 14633 & 5.259(92) & 3.271(64) & 2.049(39) & 0.519(16) & 0.209(17) & 3 & 10\\
\object{WD0839$-$327} & DA & 8930 & 37.26(65) & 24.81(48) & 16.04(30) & 4.20(13) & 1.598(52) & 3\\
\object{WD0843$+$358} & DZ & 17103 & 2.144(37) & 1.442(28) & 0.971(18) & 0.2704(100) & 0.161(24) & 5 & 7\\
\object{WD0912$+$536} & DCP & 7160 & 7.57(13) & 5.32(10) & 3.722(70) & 1.150(35) & 0.454(25) & 5\\
\object{WD1031$-$114} & DA & 25714 & 5.94(10) & 3.406(66) & 1.953(37) & 0.482(16) & 0.188(26) & 5\\
\object{WD1036$+$433} & sdO & 80000 & 35.10(61) & 24.27(47) & 16.14(30) & 4.37(13) & 1.723(57) & 5 & 4,11\\
\object{WD1053$-$550} & DA & 12099 & 2.702(47) & 1.621(32) & 1.129(21) & 0.2666(93) & 0.106(15) & 3\\
\object{WD1055$-$072} & DC & 7420 & 4.949(86) & 3.454(68) & 2.691(51) & 0.674(21) & 0.289(23) & 3 & 2\\
\object{WD1105$-$048} & DA & 16051 & 6.93(12) & 4.288(84) & 2.549(48) & 0.655(21) & 0.266(30) & 5\\
\object{WD1121$+$216} & DAZ & 7490 & 5.93(10) & 4.388(86) & 2.913(55) & 1.660(51) & 0.709(27) & 2 & 7,5\\
\object{WD1134$+$300} & DA & 20370 & 10.12(18) & 5.87(12) & 3.554(67) & 0.873(27) & 0.291(30) & 5 & 5\\
\object{WD1202$-$232} & DAZ & 8567 & 17.45(30) & 12.30(24) & 7.71(15) & 2.038(62) & 0.753(31) & 3\\
\object{WD1223$-$659} & DA & 7276 & 7.39(13) & 5.099(100) & 3.200(60) & 0.953(30) & 0.394(18) & 2 & 5\\
\object{WD1234$+$481} & DA & 55040 & 1.628(28) & 1.067(21) & 0.707(13) & 0.2116(75) & 0.085(15) & 3 & 4,12\\
\object{WD1236$-$495} & DAV & 11550 & 4.787(84) & 3.050(60) & 1.824(34) & 0.468(15) & 0.167(15) & 3\\
\object{WD1254$+$223} & DA & 40588 & 3.960(69) & 2.122(41) & 1.261(24) & 0.310(10) & 0.123(18) & 3 & 5\\
\object{WD1327$-$083} & DA & 13875 & 14.26(25) & 8.70(17) & 5.37(10) & 1.386(42) & 0.488(26) & 3 & 5\\
\object{WD1337$+$705} & DAZ & 20970 & 8.00(14) & 4.651(91) & 2.777(52) & 0.753(24) & 0.296(22) & 5\\
\object{WD1407$-$475} & DA & 18892 & 2.022(35) & 1.208(24) & 0.728(14) & 0.2093(77) & 0.082(17) & 2\\
\object{WD1408$+$323} & DA & 16465 & 2.780(48) & 1.647(32) & 1.126(21) & 0.293(10) & 0.121(21) & 5 & 5\\
\object{WD1425$-$811} & DA & 12000 & 5.91(10) & 3.708(72) & 2.322(44) & 0.594(20) & 0.234(21) & 5\\
\object{WD1509$+$322} & DA & 14371 & 2.680(47) & 1.516(30) & 0.989(19) & 0.2640(89) & 0.108(14) & 3\\
\object{WD1531$-$022} & DA & 18110 & 2.783(48) & 1.647(32) & 0.948(18) & 0.2487(85) & 0.106(17) & 3\\
\object{WD1559$+$369} & DAV & 10286 & 2.705(47) & 1.916(37) & 1.169(22) & 0.2680(89) & 0.112(13) & 2 & 9\\
\object{WD1606$+$422} & DA & 11320 & 4.063(71) & 2.511(49) & 1.599(30) & 0.409(13) & 0.165(13) & 2\\
\object{WD1611$-$084} & DA & 33214 & 2.258(39) & 1.192(23) & 0.745(14) & 0.1903(79) & 0.096(29) & 5 & 9\\
\object{WD1615$-$154} & DA & 29623 & 3.863(67) & 2.120(41) & 1.249(24) & 0.319(11) & 0.099(18) & 3\\
\object{WD1616$-$390} & DA & 11580 & 5.446(95) & 4.901(96) & 3.489(66) & 1.117(35) & 0.319(22) & 3 & 4,13\\
\object{WD1626$+$368} & DBZ & 8640 & 5.594(98) & 3.544(69) & 2.477(47) & 0.680(21) & 0.286(17) & 3\\
\object{WD1631$+$396} & DA & 20540 & 1.835(32) & 1.150(22) & 0.737(14) & 0.1651(61) & 0.057(15) & 3\\
\object{WD1637$+$335} & DA & 9940 & 2.410(42) & 1.673(33) & 1.133(21) & 0.2648(90) & 0.082(16) & 2\\
\object{WD1645$+$325} & DBV & 24600 & 3.559(62) & 2.128(42) & 1.144(22) & 0.356(12) & 0.115(21) & 5\\
\object{WD1647$+$591} & DAV & 12000 & 17.08(30) & 10.59(21) & 6.53(12) & 1.724(52) & 0.652(28) & 5\\
\object{WD1655$+$215} & DA & 9180 & 4.447(78) & 3.047(59) & 1.900(36) & 0.518(16) & 0.163(20) & 3\\
\object{WD1659$-$531} & DA & 14609 & 4.726(82) & 2.873(56) & 1.750(33) & 0.441(15) & 0.100(17) & 2\\
\object{WD1713$+$332} & DA & 20630 & 1.748(30) & 1.052(20) & 0.689(13) & 0.1882(66) & 0.064(12) & 2\\
\object{WD1716$+$020} & DA & 13470 & 2.298(40) & 1.573(31) & 0.998(19) & 0.296(10) & 0.110(18) & 2\\
\object{WD1748$+$708} & DXP & 5590 & 13.15(23) & 9.98(20) & 6.62(12) & 1.981(60) & 0.802(31) & 5\\
\object{WD1756$+$827} & DA & 7270 & 5.609(98) & 4.183(82) & 2.828(54) & 0.787(25) & 0.290(21) & 5 & 5\\
\object{WD1822$+$410} & DZ & 14350 & 2.285(40) & 1.395(27) & 0.877(17) & 0.2108(74) & 0.085(19) & 3\\
\object{WD1840$-$111} & DA & 11587 & 3.526(62) & 2.481(48) & 1.449(27) & 0.501(20) & 0.231(24) & 2 & 7\\
\object{WD1900$+$705} & DXP & 12070 & 7.39(13) & 4.312(84) & 2.865(54) & 0.720(23) & 0.309(23) & 5\\
\object{WD1919$+$145} & DA & 14838 & 7.91(14) & 4.261(83) & 2.544(48) & 0.818(34) & 0.631(73) & 2 & 7\\
\object{WD1935$+$276} & DAV & 12318 & 8.50(15) & 5.31(10) & 3.107(59) & 0.890(28) & 0.315(20) & 3 & 5\\
\object{WD1936$+$327} & DA & 18413 & 3.608(63) & 2.079(41) & 1.199(23) & 0.332(11) & 0.115(18) & 3\\
\object{WD1942$+$499} & DA & 34086 & 1.136(20) & 0.737(14) & 0.565(11) & 0.1113(50) & 0.037(13) & 3\\
\object{WD1943$+$163} & DA & 18851 & 2.550(44) & 1.441(28) & 0.936(18) & 0.2341(90) & 0.133(15) & 2 & 7\\
\object{WD1953$-$011} & DAP & 7920 & 9.43(16) & 6.29(12) & 4.153(78) & 1.259(39) & 0.505(24) & 3 & 7,2\\
\object{WD2004$-$605} & DA & 26481 & 3.845(67) & 2.193(43) & 1.384(26) & 0.334(12) & 0.154(24) & 5\\
\object{WD2007$-$303} & DA & 14990 & 14.77(26) & 8.99(18) & 5.56(10) & 1.433(44) & 0.589(34) & 5\\
\object{WD2014$-$575} & DA & 27407 & 2.936(51) & 1.667(33) & 1.040(20) & 0.2314(80) & 0.095(18) & 3\\
\object{WD2028$+$390} & DA & 31725 & 4.314(75) & 2.556(50) & 1.215(23) & 0.341(12) & 0.131(77) & 2 & 9\\
\object{WD2032$+$248} & DA & 20039 & 24.37(42) & 15.19(30) & 8.90(17) & 2.336(71) & 0.842(35) & 5\\
\object{WD2034$-$532} & DB & 13076 & 1.532(27) & 1.490(29) & 0.915(17) & 0.2409(95) & 0.091(26) & 5\\
\object{WD2039$-$202} & DA & 19373 & 11.82(21) & 6.95(14) & 4.191(79) & 1.058(33) & 0.382(23) & 3\\
\object{WD2039$-$682} & DA & 17541 & 5.139(90) & 3.075(60) & 2.013(38) & 0.499(16) & 0.187(21) & 5\\
\object{WD2046$+$396} & DA & 25296 & 1.792(31) & 1.781(35) & 1.161(22) & 0.358(12) & 0.150(21) & 2 & 7\\
\object{WD2047$+$372} & DA & 14118 & 7.61(13) & 4.595(90) & 2.826(53) & 0.774(25) & 0.224(18) & 2\\
\object{WD2105$-$820} & DAZ & 10200 & 6.48(11) & 4.265(83) & 2.575(49) & 0.736(23) & 0.295(28) & 5 & 5\\
\object{WD2111$+$498} & DA & 34386 & 5.018(87) & 3.118(61) & 1.703(32) & 0.426(14) & 0.142(26) & 2\\
\object{WD2115$+$339} & DOV & 170000 & 4.739(83) & 2.761(54) & 1.418(27) & 0.404(14) & 0.175(22) & 5 & 14\\
\object{WD2115$-$560} & DAZ & 9700 & 3.618(63) & 2.582(50) & 1.641(31) & 1.034(32) & 0.913(35) & 5 & 4,15\\
\object{WD2117$+$539} & DA & 15394 & 13.49(23) & 7.88(15) & 4.830(91) & 1.320(40) & 0.532(26) & 3 & 5\\
\object{WD2126$+$734} & DA & 14341 & 9.21(16) & 5.55(11) & 3.610(68) & 0.963(30) & 0.377(17) & 2 & 5,7\\
\object{WD2130$-$047} & DB & 17500 & 1.717(30) & 1.016(20) & 0.573(11) & 0.1477(78) & 0.073(32) & 5 & 16\\
\object{WD2134$+$218} & DA & 17814 & 1.909(33) & 1.006(20) & 0.729(14) & 0.1672(75) & \nodata & 5 & 17\\
\object{WD2136$+$828} & DA & 16400 & 6.64(12) & 4.107(80) & 2.662(50) & 0.622(20) & 0.215(20) & 5\\
\object{WD2140$+$207} & DQ & 8860 & 10.24(18) & 6.90(14) & 4.520(85) & 1.252(39) & 0.434(26) & 5\\
\object{WD2148$+$286} & DA & 60240 & 49.26(86) & 27.23(53) & 15.90(30) & 4.22(13) & 1.599(51) & 3\\
\object{WD2149$+$021} & DAZ & 17938 & 8.34(14) & 4.965(97) & 2.918(55) & 0.788(25) & 0.288(28) & 5 & 5\\
\object{WD2211$-$495} & DA & 58149 & 16.83(29) & 9.24(18) & 5.85(11) & 1.384(42) & 0.507(29) & 5\\
\object{WD2216$-$657} & DZ & 12082 & 2.439(42) & 1.617(32) & 1.031(20) & 0.2913(96) & 0.098(14) & 3\\
\object{WD2246$+$223} & DA & 10330 & 2.925(51) & 1.921(38) & 1.202(23) & 0.313(10) & 0.122(16) & 3\\
\object{WD2316$-$173} & DB & 12918 & 3.417(60) & 2.151(42) & 1.324(25) & 0.316(12) & 0.138(29) & 5\\
\object{WD2326$+$049} & DAV & 13003 & 8.91(16) & 6.03(12) & 5.60(11) & 8.80(26) & 8.64(26) & 5 & 4\\
\object{WD2329$-$291} & DAWK & 26620 & 2.451(43) & 1.481(29) & 0.895(17) & 0.2173(89) & 0.082(32) & 5\\
\object{WD2331$-$475} & DAZ & 50400 & 3.562(62) & 2.038(40) & 1.275(24) & 0.2974(99) & 0.138(16) & 3\\
\object{WD2333$-$165} & DA & 9789 & 5.83(10) & 3.344(65) & 2.181(41) & 0.558(18) & 0.207(19) & 3\\
\object{WD2359$-$434} & DAP & 8690 & 14.58(25) & 10.97(21) & 7.01(13) & 1.841(56) & 0.692(34) & 5\\
\enddata

\tablecomments{Infrared fluxes for stars in this survey. Fluxes are in units
of mJy. Values in parentheses are uncertainties for the two least significant digits for each measurment. Ap refers to the aperture size in pixels used in the photometry. Temperatures and spectral types are taken from \citet{McCook99} or Table~\ref{daz}, unless otherwise noted. JHK photometry taken from the 2MASS survey and is presented here for convenience. Notes are as follows: 
(1) Also known as GD408. Spectral type and temperature from \citet{Wolff02}.
(2) Spectral type taken from \citet{Kilic06}.
(3) Also GD619. Spectral type and temperature from \citet{Kilkenny88}.
(4) No colour correction applied to this photometry.
(5) Spectral type taken from Simbad.
(6) Temperature from \citet{Giovannini98}.
(7) Photometry of this star is contaminated by the flux from a nearby object and is not trustworthy.
(8) Photometry of this object is affected by the diffraction spike of a bright foreground object.
(9) Temperature from \citet{Finley97}.
(10) 8$\mu$m photometry slightly contaminated by background nebulosity.
(11) Spectral type and temperature from \citet{Thejll91}.
(12) Spectral type and temperature from \citet{Liebert05}.
(13) Temperature from \citet{Sion88}.
(14)  Temperature from \citet{Werner96}.
(15) Temperature from \citet{Koester06}.
(16)  Temperature from \citet{Oke84}.
(17) Object not detected at 8$\mu$m.
}
\end{deluxetable}

%% file: tab3.tex

\begin{deluxetable}{lrrll}
\tabletypesize{\footnotesize}
\tablewidth{0pt}
\tablecaption{Survey objects with detected photospheric metals \label{daz}}

\tablehead{
    \colhead{Name}&
	\colhead{[Ca/H]}&
    \colhead{$T_{eff}$}&
    \colhead{Type}&
    \colhead{Ref}\\
}

\startdata
WD0002$+$729 & $-$11.4 & 13750 & DBZ & D93\\
WD0005$+$511 & ... & 47083 & DO & H03\\
WD0455$-$282 & ... & 57273 & DA & H03\\
WD0501$+$527 & ... & 40588 & DA & H03\\
WD0552$-$041 & -10.9 & 5060 & DAZ & D93\\
WD0621$-$376 & ... & 48333 & DA & H03\\
WD0752$-$676 & -9.7 & 5730 & DAZ & D93\\
WD0843$+$358 & $-$9.6 & 17103 & DBZ & D93\\
WD1121$+$216 & $-$9.8 & 7490 & DAZ & D93\\
WD1202$-$232 & -9.8 & 8567 & DAZ & K06\\
WD1337$+$705 & $-$6.7 & 20970 & DAZ & K06\\
WD1626$+$368 & $-$8.65 & 8640 & DBZ & W02\\
WD1645$+$325 & ... & 24600 & DB & H03\\
WD1822$+$410 & $-$8.15 & 14350 & DZ & W02\\
WD2032$+$248 & ... & 20039 & DA & H03\\
WD2105$-$820 & -8.6 & 10200 & DAZ & K05\\
WD2111$+$498 & ... & 34386 & DA & H03\\
WD2115$-$560 & -7.6 & 9700 & DAZ & K06\\
WD2149$+$021 & $-$7.7 & 17938 & DAZ & K06\\
WD2216$-$657 & -9.1 & 12082 & DZ & K05\\
WD2326$+$049 & $-$6.4 & 13003 & DAZ & K06\\
WD2331$-$475 & ... & 50400 & DA & H03\\
\enddata

\tablecomments{Objects with traces of metals in their photosphere in this survey. References: D93 \citet{Dupuis93}; W02, \citet{Wolff02}; H03, \citet{Holberg03}; K05, \citet{Koester05}; K06, \citet{Koester06}
}
\end{deluxetable}